\documentclass[12pt]{article}

\usepackage{ifpdf}
\ifpdf
\usepackage{graphicx,color}
\usepackage{hyperref}
\else
\usepackage[dvipdfmx]{graphicx,color}
\usepackage[dvipdfmx]{hyperref}
\fi
\usepackage{amssymb,amsfonts,amsmath,cancel,cite,multirow}
\usepackage[capitalise]{cleveref}

\newcommand{\Ocal}[0]{\mathcal{O}}
\newcommand{\Lcal}[0]{\mathcal{L}}
\newcommand{\Wino}[0]{\widetilde W}

\newcommand{\lam}{\lambda}

\newcommand{\ep}{\epsilon}

\newcommand{\si}{\sigma}

\newcommand{\diff}[3]{
\if 1#1  \frac{\mathrm{d} #2 }{\mathrm{d} #3 }
\else  \frac{\mathrm{d}^{#1} #2 }{\mathrm{d}#3^{#1} } \fi
}
\newcommand{\ovl}{\overline}
\newcommand{\eqs}[1]{\begin{equation}\begin{split} #1 \end{split}\end{equation}}

\begin{document}

\begin{titlepage}

\begin{flushright}
	CTPU-PTC-19-03 \\
\end{flushright}

\vskip 1.35cm
\begin{center}

{\large
\textbf{
Renormalization Effects on Electric Dipole Moments in Electroweakly Interacting Massive Particle Models
}}
\vskip 1.2cm
Wataru Kuramoto$^{a}$,
Takumi Kuwahara$^{b}$, and
Ryo Nagai$^{c,d}$

\vskip 0.4cm

\textit{$^a$Department of Physics,
Nagoya University, Nagoya 464-8602, Japan
} \\
\textit{$^b$
Center for Theoretical Physics of the Universe,
Institute for Basic Science (IBS), Daejeon 34126, Korea} \\
\textit{$^c$
Institute for Cosmic Ray Research (ICRR), The University of Tokyo, Kashiwa, Chiba 277-8582, Japan
}\\
\textit{
$^d$
Department of Physics, Tohoku University, Sendai, Miyagi
980-8578, Japan
} \\

\vskip 1.5cm

\begin{abstract}

We study the renormalization effects on electric dipole moments in the models with new electroweakly interacting massive fermions.
The electric dipole moments are generated by the effective operators which arise from integrating out heavy particles at some scale in the models.
We give the renormalization group equation for the Wilson coefficients of the effective operators from the scale where the operators are generated to the electroweak scale.
Our numerical studies focus on the electric dipole moments in the mini-split supersymmetric scenario and the electroweakly interacting massive particle dark matter scenario.
It turns out that the renormalization effects can give an enhancement factor being of the order of $\Ocal(10)\%$ in the mini-split scenario and being more than two in the minimal dark matter model.

\end{abstract}

\end{center}
\end{titlepage}

\section{Introduction \label{sec:intro}}

The measurements of electric dipole moments (EDMs) of the standard model (SM) particles are powerful probes of $CP$ violation in the models beyond the standard model (BSMs).
In the SM, the EDMs are generated by three-loop diagrams for nucleons or four-loop diagrams for electrons when the $\theta$-term vanishes by a phenomenon such as Peccei-Quinn symmetry~\cite{Peccei:1977hh}, and thus EDMs are highly suppressed in the SM \cite{Gavela:1981sk,Khriplovich:1981ca,Eeg:1983mt,McKellar:1987tf,Pospelov:1991zt}.
The BSMs often provide new sources of $CP$ violation, and they can lead to large EDMs compared to the SM prediction.
Therefore, the EDMs are sensitive probes of the $CP$ violation in the new sector.

ACME collaboration has very recently reported the updated bound on the electron EDM: $|d_e| < 1.1 \times 10^{-29}~e\,\text{cm} ~ (90\% \, \text{CL})$ \cite{Andreev:2018ayy}.
This improved limit on the electron EDM is so strong that the bound allows to set constraints on the TeV-scale physics even if the new contribution to EDMs appears at two-loop order \cite{Cesarotti:2018huy,Panico:2018hal} (For earlier analysis in the framework of the SM effective field theory, see Refs.~\cite{Cirigliano:2016njn,Cirigliano:2016nyn}).
It is expected that the ACME experiment will improve the limit on electron EDM by almost two orders of magnitude at most \cite{Doyle:2016talk}.

Nucleon EDMs have a sensitivity to explore the $CP$ violating nature of the quark sector.
The current limit on proton EDM is indirectly extracted from the mercury EDM, $|d_p| < 7.9  \times 10^{-25}~e\,\text{cm}$ \cite{Griffith:2009zz}.
A proposed experiment using a storage ring will give a direct measurement and improve the sensitivity at least by four orders of magnitude compared to the current bound on proton EDM \cite{Anastassopoulos:2015ura}, $|d_p| \simeq 1.0  \times 10^{-29}~e\,\text{cm}$.
The current upper limit on the neutron EDM is given by $|d_n| < 3.0  \times 10^{-26}~e\,\text{cm} ~ (90\% \,\text{CL})$ \cite{Baker:2006ts,Afach:2015sja}.
Several proposed experiments using ultracold neutrons will achieve a sensitivity of $|d_n| \simeq 1.0  \times 10^{-28}~e\,\text{cm}$ (see review Ref.~\cite{Chupp:2017rkp}).

In this work, we consider models with a new fermion which is charged under $SU(2)_L \times U(1)_Y$ and interacts with the SM Higgs through $CP$ violating dimension-five effective operators.
We will refer to these models as electroweakly interacting massive particle (EWIMP) models.
Unless the dimension-five effective operators are highly suppressed, they provide sizable contributions to the EDMs \cite{Hisano:2014kua,Nagata:2014aoa}.
For this reason, they can be good benchmark models for the measurements of the EDMs.

In addition to the aspects of the good benchmark models, the EWIMP models have gathered attentions in several features:
the EWIMPs can be the dark matter (DM) candidate \cite{Hisano:2003ec,Hisano:2004ds,Hisano:2004pv,Hisano:2005ec,Cirelli:2005uq,Hisano:2006nn,Cirelli:2007xd,Cirelli:2009uv},
the detectability of the EWIMPs at lepton or hadron colliders has been discussed \cite{Bharucha:2018pfu,Chigusa:2018vxz,Cirelli:2014dsa,Fukuda:2017jmk,Han:2018wus,Harigaya:2015yaa,Ostdiek:2015aga,Matsumoto:2017vfu,DiLuzio:2018jwd,Matsumoto:2018ioi,Kadota:2018lrt}, and so on.

The EWIMP models can be realized as low-energy effective models of specific ultraviolet (UV) completions.
For example, in (mini-) split supersymmetric (SUSY) models \cite{ArkaniHamed:2004fb,Giudice:2004tc,ArkaniHamed:2004yi,Wells:2004di,Ibe:2006de,Hall:2011jd,Arvanitaki:2012ps,ArkaniHamed:2012gw,Ibe:2012hu}, winos and/or higgsinos, which are respectively supersymmetric partners of $W$ bosons and Higgs bosons, are the candidates of the EWIMP DM.
Therefore, the indirect and direct searches of EWIMPs have also been motivated from the point of view of model building for the UV completions.

In this paper, we study the EDMs in the EWIMP models with particular attention to the renormalization group (RG) effects.
The EDMs in the EWIMP models dominantly arise from the Barr-Zee contributions \cite{Barr:1990vd} with an insertion of the dimension-five effective operators \cite{Hisano:2014kua,Nagata:2014aoa}.
It should be noted that, in order to calculate the EDMs with sufficient accuracy, we have to evolve the Wilson coefficients of the effective operators down to the electroweak scale from the UV scale.
The RG effects are expected to become important when the UV scale is much higher than the mass scale of the EWIMP.
In this work, we derive the anomalous dimensions for the effective operators in the specific models with an EWIMP fermion: especially mini-split SUSY models with heavy higgsinos and the minimal DM models.
We obtain the precise values for the EDMs by applying our formulae, and then we discuss the impact of the RG effects on the EDMs.

This paper is organized as follows: in \cref{sec:EDM}, we briefly introduce the EWIMP models, and discuss the EDMs in the models.
Then we discuss the renormalization group equations (RGEs) above the energy scale of the mass of EWIMPs in \cref{sec:RGE}.
Numerical studies for specific models are shown in \cref{sec:numerical}; in particular, split supersymmetry scenario with heavy higgsino and minimal DM model.
\cref{sec:conclusion} is devoted to conclusions of our work.

\section{Electric Dipole Moments in EWIMP Models \label{sec:EDM}}

We consider the model with an $SU(2)_L$ $n$-plet fermion, denoted by $\chi$, with hypercharge $Y_\chi$ in addition to the SM particles.
The Lagrangian for $\chi$ being a Majorana fermion is given by
\eqs{
\Lcal_\chi & = \frac12 \left(i \ovl \chi \gamma^\mu D_\mu \chi - M \ovl \chi \chi\right)
+ \dfrac{1}{2} \tilde C_sH^\dag H \ovl \chi^C i\gamma_5 \chi \, ,
\label{eq:CPvioint_M}
}
where $D_\mu = \partial_\mu + i g T^a W^a_\mu + i g^\prime Y_\chi B_\mu$ is a gauge covariant derivative with $T^a$ being an $n$-dimensional representation of the $SU(2)_L$ generators, and with $g \,$, $g^\prime$ being the gauge coupling constants of $SU(2)_L$ and $U(1)_Y$, respectively.
$H$ denotes the SM Higgs field.
The mass parameter, $M$, is taken to be real.
$\tilde{C}_s$ denotes the Wilson coefficient of the effective interactions.
The Lagrangian for $\chi$ being a Dirac fermion, on the other hand, is given by \footnote{
For the case of the doublet EWIMP fermion, the effective operator, such as $(\ep \ovl \chi^C H)(\ep \chi H)$ with the $SU(2)$ antisymmetric tensor $\ep$, is also allowed.
This operator, however, does not induce the EDMs via the Barr-Zee diagrams.
The phenomenology of these operators has been intensively discussed in the higgsino dark matter model in a split SUSY scenario \cite{Nagata:2014wma}, and the RGEs of these operators have been derived in the context of neutrino mass operators \cite{Babu:1993qv,Chankowski:1993tx}.
}
\eqs{
\Lcal_\chi & = i \ovl \chi \gamma^\mu D_\mu \chi - M \ovl \chi \chi
+ \tilde C_s H^\dag H \ovl \chi i\gamma_5 \chi
+ \tilde C_t H^\dag t^a H \ovl \chi i\gamma_5 T^a \chi \, ,
\label{eq:CPvioint_D}
}
where $t^a$ is a fundamental representation of the $SU(2)_L$ generators.
$\tilde{C}_s$ and $\tilde{C}_t$ denote the Wilson coefficients of the effective interactions.

We note that the Wilson coefficients are determined in terms of fundamental couplings and masses of heavy particles once we specify a UV completion.
A concrete example for the UV completion is discussed in \cref{sec:heavyhiggsino}.
At this stage, we do not specify the detail of the UV completion.

The EWIMP fermions can have the dipole-type interactions such as $\ovl \chi \si_{\mu\nu} \gamma_5 T^a \chi W^{a \mu\nu}$ with the W-boson field strength tensor $W^{a \mu\nu}$.
However, these operators arise at the one-loop level and they have an additional suppression factor due to the chirality property of the dipole operators.
Therefore, we expect the dipole operators of the EWIMP fermions to be subdominant in this study.
We also assume that the $CP$-odd dimension-six operators of the SM fields are negligible.
This assumption is valid when the UV scale is much higher than the mass scale of the EWIMP mass and/or the Wilson coefficients of the dimension-six operators are accidentally suppressed.
We will disscuss the UV completion which satisfies this assumption in \cref{sec:heavyhiggsino}

In this model, the $CP$-odd effective interactions induce the EDMs for light fermions via two-loop diagrams, so-called Barr-Zee diagrams~\cite{Barr:1990vd}.
The EDM for a light fermion $f$ with mass $m_f$ and an electromagnetic charge $Q_f$ is computed as follows \cite{Hisano:2014kua,Nagata:2014aoa}.
\eqs{
d_f = d_f^{\gamma H} + d_f^{ZH} + d_f^{WW} \,.
\label{eq:EDM_f}
}
For $\chi$ being a Majorana fermion, each of the contributions is calculated as
\eqs{
\frac{d_f^{\gamma H}}{e} & = \frac{\alpha_e m_f Q_f}{6 (4\pi)^3}
\frac{n(n^2 - 1)}{M} \tilde C_s f_0\left( \frac{M^2}{m_h^2} \right) \,, \\
\frac{d_f^{ZH}}{e} & = \frac{\alpha m_f}{12 (4\pi)^3} (T^3_f - 2 Q_f \sin^2 \theta_W)
\frac{n(n^2 - 1)}{M} \tilde C_s f_1 \left(\frac{m_Z^2}{m_h^2}, \frac{M^2}{m_h^2}\right)\,, \\
\frac{d_f^{WW}}{e} & = 0 \,.
}
Note that $d^{WW}_f$ vanishes due to the cancellation between the contributions from the positive-charged fermions and the negative-charged fermions.
Similarly, for $\chi$ being a Dirac fermion, it is calculated as
\eqs{
\frac{d_f^{\gamma H}}{e} & = \frac{\alpha_e m_f Q_f}{3 (4\pi)^3} \frac{n}{M} f_0\left( \frac{M^2}{m_h^2} \right)
\left[ (n^2 - 1 + 12 Y_\chi^2) \tilde C_s - Y_\chi (n^2-1) \tilde C_t \right]\,, \\
\frac{d_f^{ZH}}{e} & = \frac{\alpha m_f}{12 (4\pi)^3} \frac{n}{M} (T^3_f - 2 Q_f \sin^2 \theta_W) f_1\left(\frac{m_Z^2}{m_h^2}, \frac{M^2}{m_h^2}\right) \\
& ~ \times \left[ 2 (n^2 - 1 - 12 Y_\chi^2 \tan^2\theta_W) \tilde C_s - Y_\chi (n^2-1) (1 - \tan^2 \theta_W) \tilde C_t \right] \,, \\
\frac{d_f^{WW}}{e} & = \frac{\alpha m_f T^3_f}{6 (4\pi)^3} \frac{n}{M} f_0\left( \frac{M^2}{m_W^2} \right) Y_\chi (n^2-1)  \tilde C_t\,.
}
Here, $\alpha_e$ and $\alpha$ are the fine structure constants of the electromagnetic interaction and of the $SU(2)_L$ one, respectively.
$T_f^3$ is the isospin of the fermion $f$.
$m_Z\,,m_W$, and $m_h$ are the masses of the $Z$ boson, $W$ boson, and Higgs boson, respectively.
$f_0(x)$ and $f_1(x,y)$ are loop functions defined by
\eqs{
f_0(r) & = r \int^1_0 dx \frac{1}{x(1-x)-r} \ln \frac{x(1-x)}{r} \,, \\
f_1(r,s) & = \frac{1}{1-r} \left[ f_0(s) - r f_0(s/r) \right] \,.
}

In our numerical study, we also focus on the nucleon EDMs, namely the proton EDM and the neutron EDM.
We evaluate the nucleon EDMs from the quark EDMs which are computed using \cref{eq:EDM_f} at the electroweak scale.
The one-loop QCD corrections \cite{Degrassi:2005zd} are taken into account between the electroweak scale and the hadronic scale, and then we estimate the nucleon EDMs at the hadronic scale, 1~GeV, using the following relations derived in Refs.~\cite{Hisano:2012sc,Hisano:2015rna}:
\eqs{
d_p & = 0.78 d_u - 0.20 d_d + e( - 1.2 \widetilde d_u - 0.15 \widetilde d_d) \, \\
d_n & = - 0.20 d_u + 0.78 d_d + e(0.29 \widetilde d_u + 0.59 \widetilde d_d) \, .
}
where $\widetilde d_q \,, ~~ (q = u, d)$ are quark chromoelectric dipole moments.
Note that we use the relations obtained when the Peccei-Quinn symmetry is assumed; that is, the QCD $\theta$ term vanishes.~\footnote{
While we use the relation based on the QCD sum rule and the lattice QCD simulation of low-energy constants, the strange chromo-EDM contribution $\widetilde d_s$ appears using (baryon) chiral perturbation theory \cite{Fuyuto:2012yf,Engel:2013lsa} even if we assume the Peccei-Quinn symmetry.
See also Ref.~\cite{Dekens:2018bci} for the expression of nucleon EDMs in the chiral perturbation approach.
In our case, the quark chromo-EDMs are generated only from the RGE mixing below the electroweak scale, and are not induced from the two-loop matching contribution since the EWIMP fermion has no color charge.
}

\section{Renormalization Group Equations \label{sec:RGE}}

As we discussed in the previous section, the $CP$ violating effective interactions given by \cref{eq:CPvioint_M,eq:CPvioint_D} induce the EDMs of the SM fermions.
When the effective operators are generated at a very high energy scale, we should evolve the Wilson coefficients to the electroweak scale.
For that reason, we will derive the RGEs for the Wilson coefficients of the $CP$ violating dimension-five effective operators in this section.

The RGEs for the Wilson coefficients are written as
\eqs{
\diff{1}{}{\ln \mu}
\left(
\begin{array}{ccc}
\tilde C_s(\mu)\\
\tilde C_t(\mu)\\
\end{array}
\right)
=
\left(
\begin{array}{ccc}
(\gamma_{\Ocal})_{ss} & (\gamma_{\Ocal})_{st}\\
(\gamma_{\Ocal})_{ts} & (\gamma_{\Ocal})_{tt}\\
\end{array}
\right)^T
\left(
\begin{array}{ccc}
\tilde C_s(\mu)\\
\tilde C_t(\mu)\\
\end{array}
\right)
\label{eq:RGE_Wilson}
}
with $\mu$ the renormalization scale.
$\gamma_{\Ocal}$ denotes the anomalous dimension matrix for the effective operators.
In this work, we compute the anomalous dimensions $\gamma_{\Ocal}$ at the one-loop level.~\footnote{
We choose the Feynman--'t Hooft gauge throughout this paper.
}
We here focus only on the corrections from the gauge interactions, the top Yukawa coupling $y_t$, and the Higgs quartic coupling $\lam$.
Indeed, these corrections are dominant when Yukawa interactions of the EWIMPs to the SM Higgs boson are negligible.~\footnote{
For instance, if we introduce a wino-like $(\mathbf{3},0)$ triplet, we can write the Yukawa interaction among the SM Higgs, the lepton doublet, and the triplet.
If we introduce two fermionic EWIMPs; wino-like $(\mathbf{3},0)$ and higgsino-like $(\mathbf{2},1/2)$ under $SU(2)_L \times U(1)_Y$, they can also have a Yukawa interaction with the SM Higgs.
If these couplings are larger than the gauge couplings, the dominant contribution can be altered.}

\begin{figure}
	\centering
	\includegraphics[width=10cm,clip]{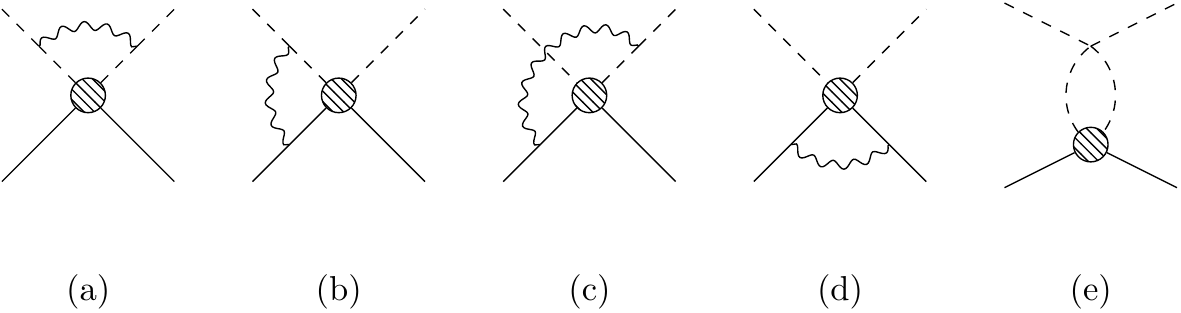}
	\caption{
	One-loop diagrams with the effective interactions.
	The blobs denote the effective vertex.
	The solid, dashed, and wavy lines correspond to the propagators of the EWIMP fermion, the Higgs field, and the SM weak gauge bosons, respectively.
	}
	\label{fig:oneloop}
\end{figure}

The relevant Feynman diagrams for computing $\gamma_{\Ocal}$ are shown in \cref{fig:oneloop}.
The blobs in the diagrams indicate the insertion of the effective operators.
We also include the diagrams where the gauge boson propagator attaches to the opposite side external lines in \cref{fig:oneloop} (b) and (c).
It turns out that the anomalous dimension matrix at one-loop level is given as~\footnote{
We found that the anomalous dimension for the effective operators was discussed in Ref.~\cite{Bishara:2018vix} in the context of the DM effective interactions.
The anomalous dimension matrix that we derived is consistent with their work.

For the $CP$-even dimension-five operators among the EWIMP fermions and the SM Higgs, the anomalous dimension matrix is the same as \cref{eq:anomalousdim}.
}
\eqs{
(\gamma_{\Ocal})_{ss} & = - \frac{1}{(4 \pi)^2} \left[ 6 g^2 \left( C_2(H) + C_2(\chi) \right) + 6 g^{\prime 2} \left( Y_H^2 + Y_\chi^2 \right) - 3 \lam - 6 y_t^2 \right] \,, \\
(\gamma_{\Ocal})_{tt} & = - \frac{1}{(4 \pi)^2} \left[ 6 g^2 \left(C_2(H) + C_2(\chi) - \frac12 C_2(G) \right) + 6 g^{\prime 2} \left( Y_H^2 + Y_\chi^2 \right) - \lam - 6 y_t^2 \right] \,, \\
(\gamma_{\Ocal})_{st} & = (\gamma_{\Ocal})_{ts} = 0 \,,
\label{eq:anomalousdim}
}
where $C_2$ is the quadratic Casimir invariant for $SU(2)_L$ representations, $t^a t^a = C_2(H) \mathbf{1}$, $T^a T^a = C_2(\chi) \mathbf{1}$, and $C_2(G)$ for an adjoint representation.
$Y_H = 1/2$ is the hypercharge of the SM Higgs.
The diagram (e) shows a contribution from the Higgs quartic coupling defined as
\eqs{
\Lcal_H = - \frac{\lam}{2} (H^\dag H)^2 \,.
}
The diagrams (b) and (c) can give the operator mixing in general, however these contributions cancel due to the gauge interaction of the Higgs field, $i g (-\partial_\mu H^\dag t^a H +  H^\dag t^a \partial_\mu H) W^{a\mu}$, at this loop level.
Therefore, the operators do not mix with each other during the one-loop RGE evolution.

This anomalous dimension implies that the Wilson coefficients tend to be enhanced during RGE evolution to the electroweak scale when a large representation for the EWIMP $\chi$ is assumed.
For a Dirac fermion $\chi$, $\tilde C_t$ also develops during RGE evolution but the renormalization factor for $\tilde C_t$ is smaller than that for $\tilde C_s$ because of the smaller coefficients in $(\gamma_{\Ocal})_{tt}$.

We note that the resultant EDMs can be more affected by the $CP$-odd dimension-six operators when the input scale is close to the electroweak scale.
Indeed, in that case, the RGE evolution becomes less important and the $CP$-odd dimension-six operators get more important for computing the EDMs.

\section{Numerical Analysis \label{sec:numerical}}
In this section, we perform the numerical analysis of the EDMs induced by the $CP$-odd dimension five effective operators and discuss the impact of the RG effects on the EDMs.
We consider effective models with a single EWIMP fermion in \cref{sec:MDM} at first, and then we will provide a discussion on the RG effect in the specific model, the split supersymmetric model with heavy higgsinos, in \cref{sec:heavyhiggsino}.

\subsection{Electroweakly Interacting Dark Matter Models \label{sec:MDM}}
First, we discuss the models where a single EWIMP fermion is introduced to examine the impact of the RG effects of the effective operators on the EDMs.
The following fermions are introduced:
a $(n,Y_\chi)=(5,0)$ fermion and a $(n,Y_\chi)=(3,0)$ fermion for Majorana EWIMP models, and a $(n,Y_\chi)=(4,1/2)$ fermion and a $(n,Y_\chi)=(2,1/2)$ fermion for Dirac EWIMP models.
Each representation contains the electromagnetically neutral state which can be a DM candidate \cite{Cirelli:2005uq,Cirelli:2007xd,Cirelli:2009uv}.~
\footnote{
The direct searches of the Dirac EWIMP DM give severe constraints on the EWIMP models due to the coupling to $Z$ boson.
However, it can be relaxed if there are additional singlet Majorana fermions which mix with the neutral component of the EWIMPs \cite{Cohen:2011ec,Cirelli:2014dsa,Bharucha:2018pfu}.
These singlets do not affect the prediction of the EDMs.
}
For instance, the 5-plet Majorana fermion is a good candidate for the EWIMP DM since the stability is automatically ensured due to the absence of decay operators, and therefore the 5-plet Majorana fermion is now referred to as the minimal dark matter (MDM).
The preferred mass for the thermal abundance of the MDM is $\Ocal(10)~\mathrm{TeV}$ \cite{Cirelli:2009uv} with the assumption that the DM dominantly consists of the MDM.
However, since the gamma ray test for the MDM excludes the 5-plet MDM with the mass of $\Ocal(10)~\mathrm{TeV}$ \cite{Cirelli:2015bda,Garcia-Cely:2015dda}, several extensions for the 5-plet MDM are proposed in Ref.~\cite{DelNobile:2015bqo}.
Furthermore, the direct detection of the MDM may also be difficult due to the required mass for the thermal abundance \cite{Hisano:2011cs}.

Therefore, in the following analysis, we do not assume that the whole DM is composed by the EWIMP, and then the lighter EWIMP DM is still viable.
We can get the direct constraint on the mass of a 5-plet at the LHC similar to that on the wino mass \cite{Ostdiek:2015aga}; it is na\"ively expected that the mass below about 400~GeV is excluded at $\sqrt{s} = 13~\mathrm{TeV}$ from the exclusion limit of the wino mass \cite{Aaboud:2017mpt}.
The EWIMP DM scenarios are expected as good benchmark models for DM physics, and we expect that the EDM can play a role in complementary searches for them.
Note again that we consider not only the 5-plet fermion but also the other EWIMP fermions.

In general, the UV-physics effects induce the effective interactions between the EWIMP DM and the SM Higgs fields \cite{Hisano:2014kua,Nagata:2014aoa}.
The effective interactions are given by \cref{eq:CPvioint_M,eq:CPvioint_D}.
As we discussed in \cref{sec:EDM}, the effective interactions induce the EDMs of the SM fermion through the Barr-Zee diagrams.
The EDMs induced by the interactions depend on the Wilson coefficients $\tilde{C}_{s,t}(m_Z)$, the mass of the EWIMP $\chi$, and the representation of $\chi$.

We parameterize the values of $\tilde{C}_{s,t}$ at the UV scale $M_{\mathrm{phys}}$ as
\eqs{
\tilde{C}_i(M_{\mathrm{phys}})
=
\frac{\xi_i}{M_{\mathrm{phys}}}
~~~~(i=s,t),
}
with $\xi_s$ and $\xi_t$ being dimensionless parameters.
Once we specify the UV completion of the EWIMP DM scenarios, we can determine $\xi_s$ and $\xi_t$ in terms of the fundamental parameters in the UV physics.
However, we do not specify the UV completion, and therefore we treat $\xi_s$ and $\xi_t$ as free parameters in the following analysis.

\begin{figure}
	\centering
  \includegraphics[width=7cm,clip]{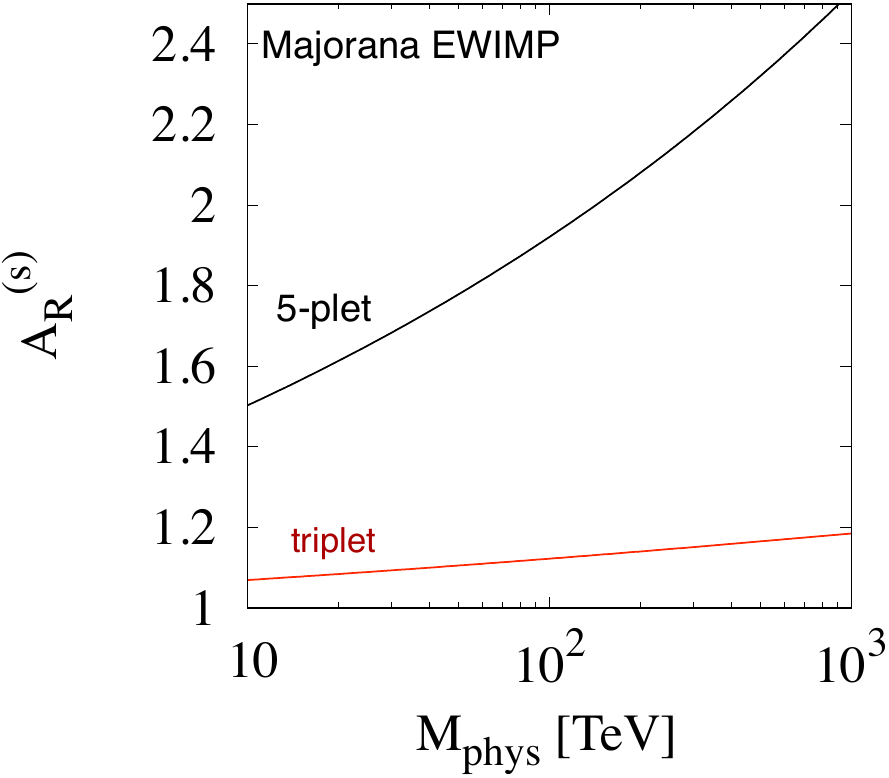}
	\includegraphics[width=7cm,clip]{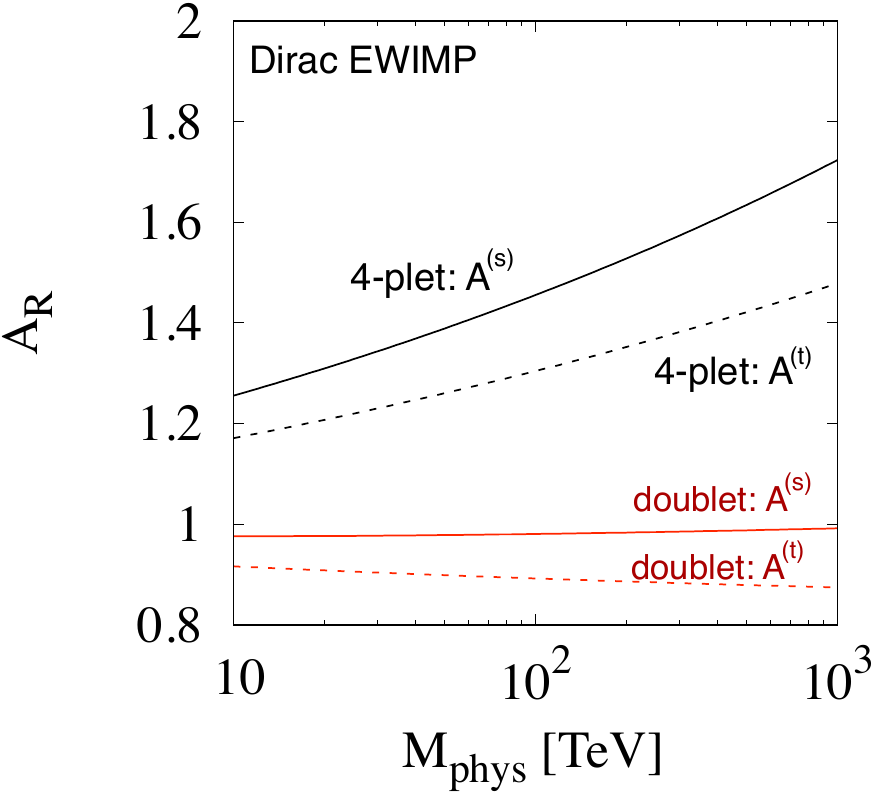}
	\caption{
	The renormalization factors in the EWIMP DM scenarios.
	We fix $\xi_s=\xi_t=0.1~(\xi_s=0.1)$ for the Dirac (Majorana) EWIMP models.
	(Left) $A^{(s)}_R$ in the model with Majorana EWIMPs: 5-plet (black) and triplet (red).
	(Right) solid (broken) lines show $A^{(s)}_R$ ($A^{(t)}_R$) in the model with Dirac EWIMPs : 4-plet (black) and doublet (red).
	}
	\label{fig:AR_ewimp}
\end{figure}

Let us first discuss the impact of the renormalization factors $A^{(s)}_R=\tilde{C}_s(m_Z)/\tilde{C}_s(M_{\mathrm{phys}})$ and $A^{(t)}_R=\tilde{C}_t(m_Z)/\tilde{C}_t(M_{\mathrm{phys}})$ in each EWIMP DM scenario.
We use the one-loop beta functions in the SM plus the EWIMP DM above the energy scale of $m_Z$ in order to compute the SM gauge couplings at the UV scale.
\cref{fig:AR_ewimp} shows the renormalization factors as the function of the UV scale $M_\mathrm{phys}$.
In the analysis, we assume $\xi_s=\xi_t=0.1~(\xi_s=0.1)$ for the Dirac (Majorana) EWIMP models.
While the red lines show a doublet or a triplet in each fermion case, the black lines show the higher representations in each fermion case: a 5-plet for the Majorana case and a 4-plet for the Dirac case.
We show $A^{(s)}_R$ in solid lines and $A^{(t)}_R$ for the Dirac EWIMP models.
We find that the electron EDM induced by the effective operators with the higher representation EWIMPs suffers from the larger RG effects.
For Dirac EWIMP cases, $A^{(t)}_R$ is less enhanced than $A^{(s)}_R$ since the anomalous dimension for $\tilde C_t$ is smaller than that for $\tilde C_s$ [see \cref{eq:anomalousdim}].
As reference values for the enhancement factor, we take $M=1~\mathrm{TeV}$ and $M_{\mathrm{phys}}=100~\mathrm{TeV}$, and then we obtain $A^{(s)}_R \simeq 1.91$ for the 5-plet EWIMP model and $A^{(s)}_R\simeq 1.00$ and $A^{(t)}_R\simeq 0.85$ in the model with a doublet EWIMP.

\begin{figure}
	\centering
	\includegraphics[width=7cm,clip]{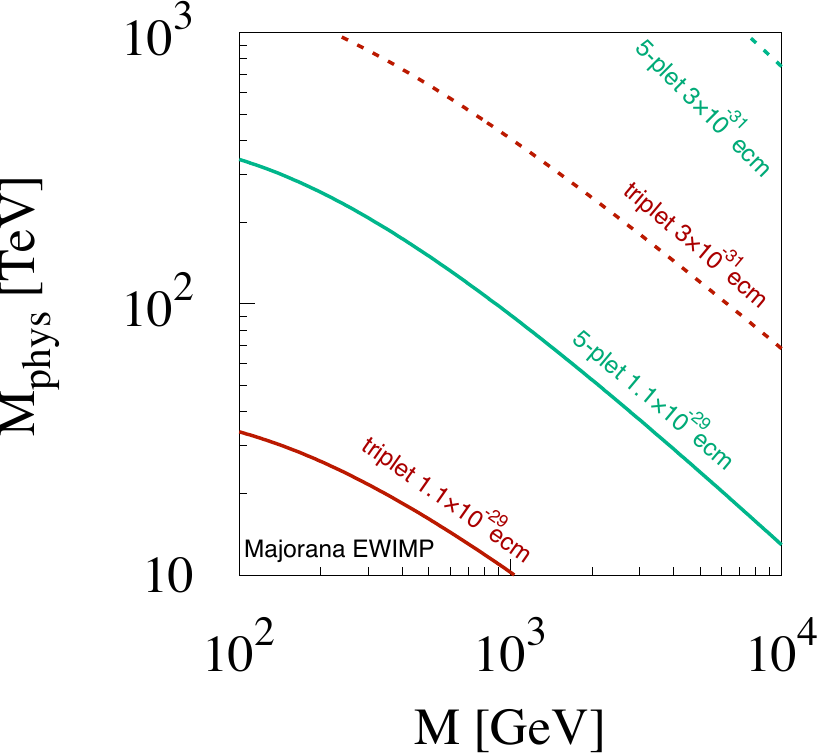}
	\includegraphics[width=7cm,clip]{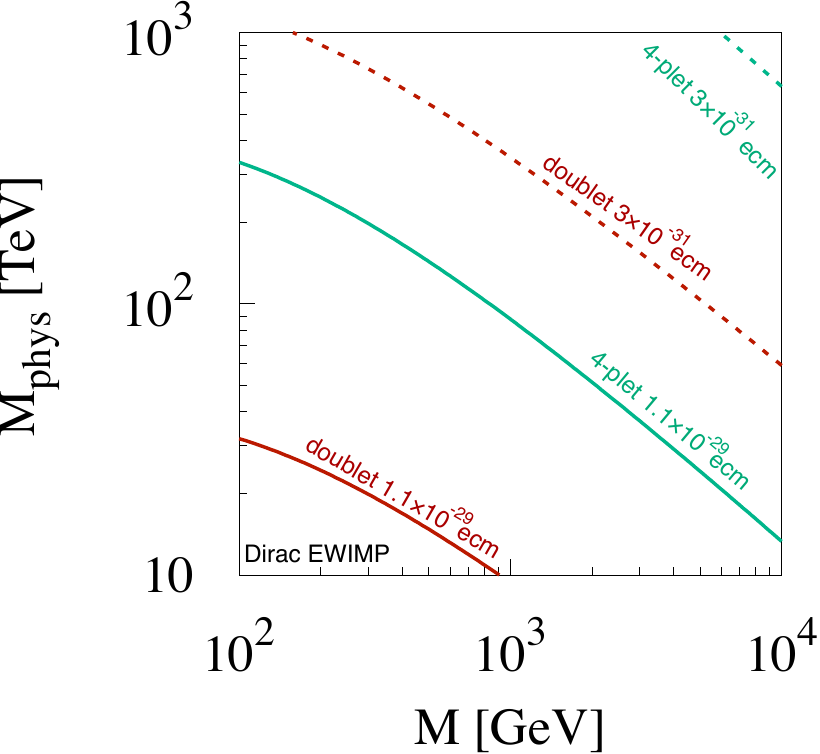}
	\caption{
	The electron EDM ($d_e$) in the EWIMP DM scenarios. We again assume $\xi_s=\xi_t=0.1~(\xi_s=0.1)$ for the Dirac (Majorana) EWIMP models.
	The solid and dotted lines corresponds to the current constraint ($|d_e|=1.1\times 10^{-29} e~\mathrm{cm}$) and the projected ACME-III sensitivity ($|d_e|=3\times 10^{-31} e~\mathrm{cm}$), respectively.
	(Left) $d_e$ in the model with a 5-plet (green) and triplet (red) Majorana fermion.
	(Right) $d_e$ in the model with a 4-plet (green) and doublet (red) Dirac fermion.
	}
	\label{fig:eEDM_ewimp}
\end{figure}

Lastly \cref{fig:eEDM_ewimp} shows the electron EDMs in the EWIMP DM models with the RG effect.
We here take $\xi_s=\xi_t=0.1~(\xi_s=0.1)$ for the Dirac (Majorana) EWIMP models.
In the figures, the differences of colors indicate the differences of the representations.
The solid and dotted lines correspond to the parameter spaces where the predicted value of electron EDM equals the current constraint ($|d_e|=1.1\times 10^{-29} e~\mathrm{cm}$) and the projected ACME-III sensitivity ($|d_e|=3\times 10^{-31} e~\mathrm{cm}$), respectively.
Since the EDMs from the Barr-Zee diagrams increase according to $n^3$, it is expected that, for large representations with $\xi_s = \xi_t = 0.1$, the broad region in the $M$-$M_\mathrm{phys}$ plane can be explored in future experiments.

\subsection{Split Supersymmetry with Heavy Higgsino \label{sec:heavyhiggsino}}
In this subsection, we focus on the mini-split SUSY spectrum, where scalar fields except the SM Higgs get masses of the order of $10^2$-$10^3$~TeV while the masses of gauginos are in the range of $10^2$~GeV to TeV.
The mini-split spectrum has attractive features.
First, several constraints on the mini-split SUSY models are milder than those on TeV-scale SUSY models: SUSY-flavor and SUSY-$CP$ constraints (for instance, see review Ref.~\cite{Gabbiani:1996hi}) thanks to the heavy sfermions, and cosmological problems (such as gravitino problems \cite{Kawasaki:2004qu,Kawasaki:2008qe,Kawasaki:2017bqm}).
Second, the minimal $SU(5)$ grand unification model is still viable if the mini-split spectrum is assumed: coupling unification \cite{ArkaniHamed:2004fb,Giudice:2004tc,ArkaniHamed:2012gw,Hisano:2013cqa} and proton decay constraints \cite{Liu:2013ula,Hisano:2013exa,Nagata:2013sba,Hall:2014vga,Evans:2015bxa,Bajc:2015ita,Ellis:2015rya}.

Let us consider the case where higgsinos are much heavier than gauginos.
This sparticle spectrum has been studied in the literature \cite{Ibe:2006de,Hall:2011jd,ArkaniHamed:2012gw,Arvanitaki:2012ps,Hall:2012zp,Ibe:2012hu,Nomura:2014asa}.
Even if we assume the heavy spectrum for sfermions and higgsinos, there are several contributions to the EDMs: for instance, there can be $CP$ violating four-Fermi interactions among quarks and gluinos \cite{Hisano:2008hn,Altmannshofer:2013lfa,Fuyuto:2013gla} and there can also be one-loop contributions when higgsinos are as heavy as sfermions \cite{Hisano:2008hn,Altmannshofer:2013lfa,McKeen:2013dma}.
However, those contributions are quickly suppressed when the flavor violations in the sfermion sector are suppressed, and therefore it is reasonable to consider the EDM only from the Barr-Zee contributions in the mini-split spectrum \cite{Giudice:2005rz}.

The neutral wino is the lightest supersymmetric particle (LSP) when the anomaly mediated spectrum \cite{Giudice:1998xp,Randall:1998uk} is assumed.
The search for the disappearing track signature at ATLAS excludes the pure wino LSP below a mass of $460$~GeV \cite{Aaboud:2017mpt}.
The thermal relic abundance of the wino LSP requires about 3~TeV for the LSP mass \cite{Hisano:2006nn,Cirelli:2007xd}.
In this study, since higgsinos are as heavy as sfermions, the bino and the wino get threshold corrections from higgsinos \cite{Giudice:1998xp}.
It is also possible that there are threshold corrections from the extra matter with mass of the same order of the typical sfermion scale $M_S$ \cite{Harigaya:2013asa}.
Therefore, we do not restrict the gaugino mass spectrum.
Although the bino mass is irrelevant for our study, it is important for the direct search of the LSP; for instance, the compressed bino-wino system where the correct abundance of the bino LSP is given \cite{ArkaniHamed:2006mb,Baer:2005jq,Ibe:2013pua,Harigaya:2014dwa} changes the collider signature \cite{Han:2014xoa,Nagata:2015pra,Duan:2018rls}.

Below the energy of the mass of higgsino $\mu_H$, we obtain the following effective interactions,~
\footnote{
At the mass of higgsino $\mu_H$, we also obtain the effective interaction between bino and the SM Higgs field.
However, the effective interaction does not make any contributions to the EDMs for light fermions.
}
\eqs{
\Lcal_{\mathrm{int}} = \frac{1}{2}\widetilde C_{\Wino} H^\dag H
\ovl{\Wino^a} i \gamma_5 \Wino^a\, ,
\label{eq:susyeffint}
}
with $\Wino$ denoting four-component Majorana fermions of the wino.
It should be noted that this interaction is nothing but the special case with a Majorana fermion with $n = 3$ and $Y_\chi = 0$ of the model discussed in \cref{sec:EDM}.

The Wilson coefficient $\widetilde C_{\Wino}$ satisfies the following matching condition at the energy scale of $\mu_H$.
\eqs{
\widetilde C_{\Wino}(\mu_H) = \frac{g_u g_d}{\mu_H} \sin \eta \, ,
\label{eq:CatmuH}
}
with $g_u$ and $g_d$ denoting gaugino couplings which satisfy $g_u = g \sin \beta$ and $g_d = g \cos \beta$ at the sfermion mass scale.
$\tan \beta = v_u/v_d$ is the ratio of the vacuum expectation values of the two Higgs doublets in the MSSM, and we assume $\tan \beta = 3$ which is a typical value for the mini-split SUSY spectrum.
$\eta$ collectively denotes the $CP$ phase in the UV theory, $\eta = \arg (g_u^\ast g_d^\ast M_2 \mu_H)$ with $M_2$ being the mass of the wino.

\begin{figure}
	\centering
	\includegraphics[width=7cm,clip]{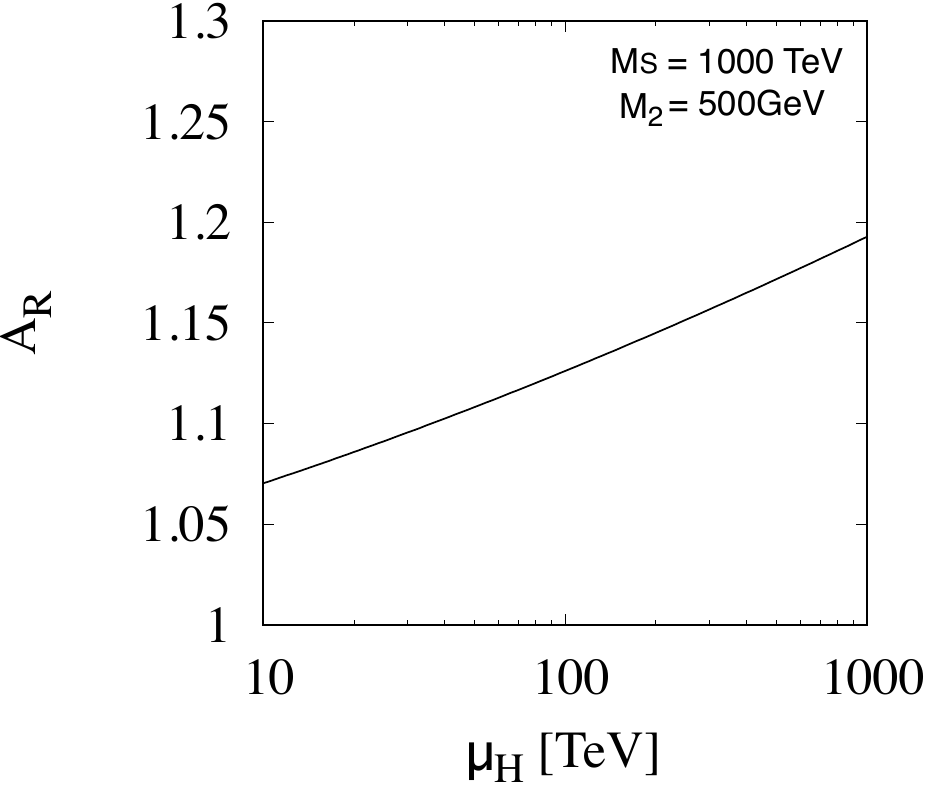}
	\caption{
	The renormalization factor as a function of $\mu_H$ in the mini-split SUSY scenario.
	$\sin\eta=1$ and $M_S=10^3~\mathrm{TeV}$ are assumed.
	We take $M_2=500~\mathrm{GeV}$.
	}
	\label{fig:ARsusy}
\end{figure}

In the following analysis, we compute the parton level EDMs as follows.
We compute the gaugino couplings $g_u$ and $g_d$ at the sfermion scale $M_S$ using the two-loop RGEs for the gauge couplings and the one-loop RGEs for the Yukawa couplings.
We note that we use the beta functions in the SM plus gauginos from the energy scale of $m_Z$ to the energy scale of $\mu_H$.
We then get $\widetilde C_{\Wino}(\mu_H)$ using \cref{eq:CatmuH} while evolving the gaugino couplings according to the one-loop RGE for them \cite{Giudice:2004tc} if $\mu_H < M_S$.
We obtain $\widetilde C_{\Wino}(m_Z)$ by taking into account the RG evolution down to the energy scale of $m_Z$.
Then, we evaluate the EDMs using \cref{eq:EDM_f} at the energy scale of $m_Z$.

First, we show the renormalization factor, $A_R=\widetilde C_{\Wino}(m_Z)/\widetilde C_{\Wino}(\mu_H)$, in \cref{fig:ARsusy} to see the impact of the RG effects on the electron EDM.
As we expected, we easily find that a larger UV scale $\mu_H$ gives a larger renormalization factor $A_R$.
We take $M_2=500$~GeV, $\sin\eta=1$, and the mass of the sfermion to be $M_S = 10^3$~TeV in this figure; however, $A_R$ does not depend on them, especially $M_S$ unless $M_S \leq \mu_H$.
As a reference value, we obtain $A_R\simeq 1.19$ when we take $\mu_H=10^3$~TeV.

\begin{figure}
	\centering
	\includegraphics[width=7cm,clip]{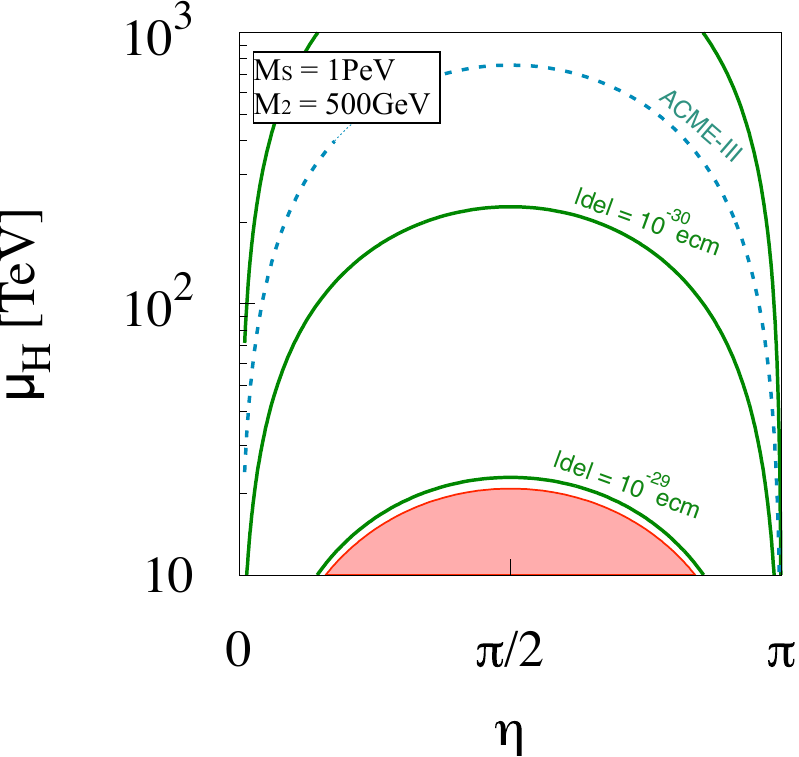}
	\includegraphics[width=7cm,clip]{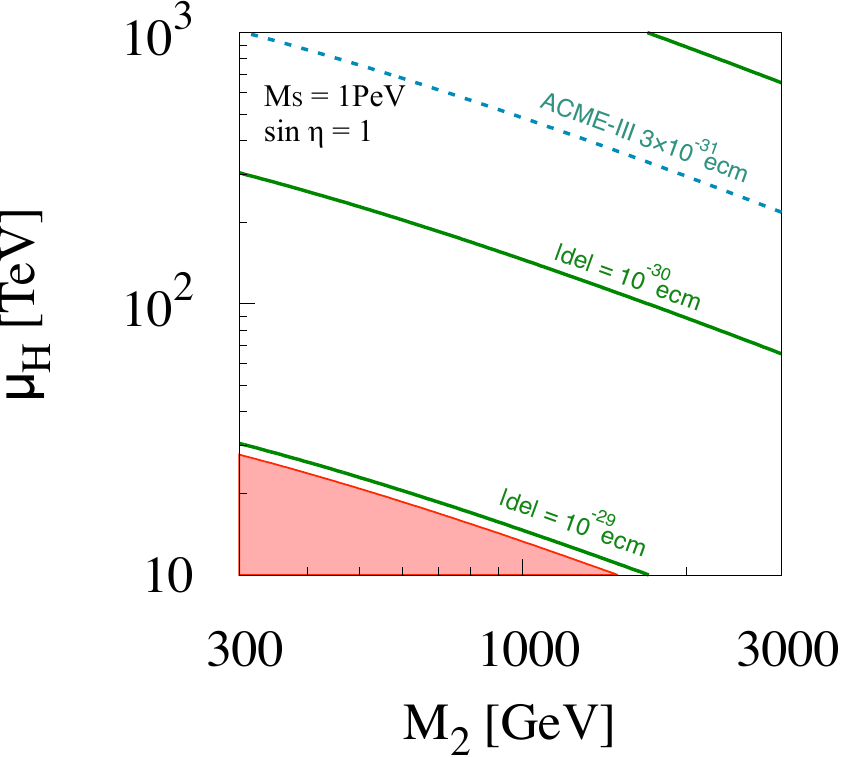}
	\caption{
	Parameter dependences of $d_e$ with $M_S = 1~\mathrm{PeV}$.
	The red shaded region is excluded by the first result of $d_e$ measurement at ACME-II.
	The dotted line shows the future ACME-III sensitivity in both panels.
	The left panel shows $\eta$-$\mu_H$ dependence of $d_e$ for fixed $M_2 = 500~\mathrm{GeV}$, while the right panel shows $M_2$-$\mu_H$ dependence of $d_e$ for fixed $\sin\eta = 1$.
	}
	\label{fig:eEDM_splitSUSY}
\end{figure}

In \cref{fig:eEDM_splitSUSY}, we show the parameter dependences of $d_e$ including the RG effects. We here assume that the mass of sfermions is $M_S = 1~\mathrm{PeV}$.
The right panel shows the $M_2$-$\mu_H$ dependence with $\sin\eta = 1$, whereas the left panel shows the $\eta$-$\mu_H$ dependence with $M_2=500~\mathrm{GeV}$.
The solid lines corresponds to $d_e = 10^{-29} \,, 10^{-30} \,,$ and $10^{-31}~e\,\mathrm{cm}$ from bottom to top in these panels.
The red shaded region is excluded by the first result of ACME-II \cite{Andreev:2018ayy}; $|d_e| \leq 1.1 \times 10^{-29}~e\,\mathrm{cm}$.
The dotted line shows the final goal of ACME-III, $|d_e| \sim 3 \times 10^{-31}~e\,\mathrm{cm}$ \cite{Doyle:2016talk}.
We find that the broad parameter region will be explored by the future sensitivity of the ACME experiment.

\begin{figure}
	\centering
	\includegraphics[width=7cm,clip]{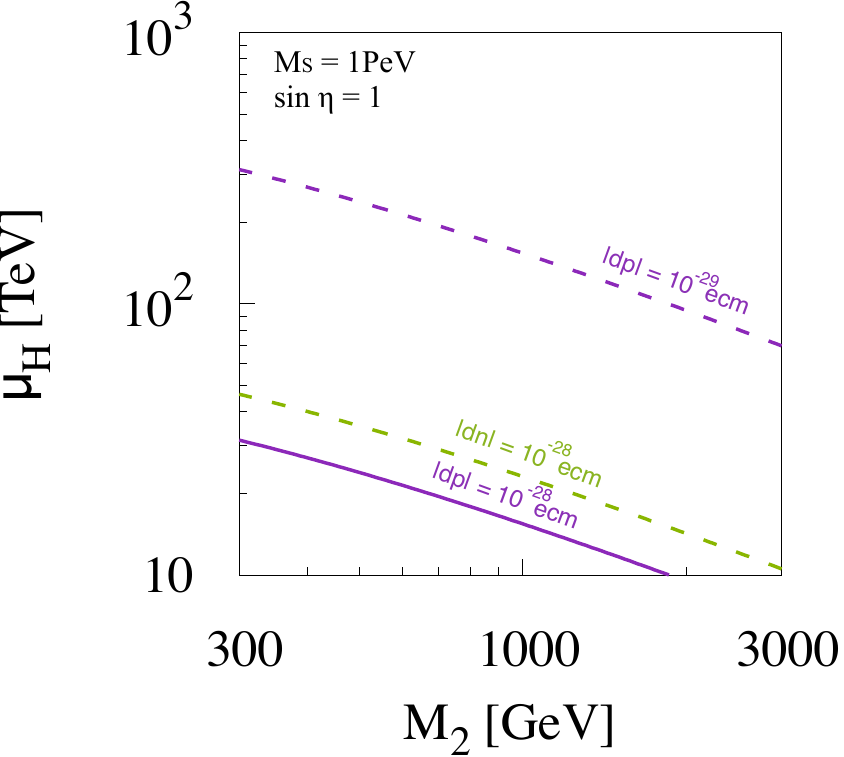}
	\caption{
	Parameter dependence of the nucleon EDMs, i.e., that of the proton EDM (purple lines) and that of the neutron EDM (green lines).
	$M_S = 1~\mathrm{PeV}$ and $\sin\eta = 1$ are assumed.
	The broken lines show the future sensitivities of nucleon EDMs.
	}
	\label{fig:NEDM_splitSUSY}
\end{figure}

The future experiments for the nucleon EDMs will also have good sensitivities.
In \cref{fig:NEDM_splitSUSY}, we show nucleon EDMs arising from the effective operators in the mini-split SUSY model.
We assume $M_S = 1~\mathrm{PeV}$ and $\sin \eta = 1$ in this figure, again.
Purple lines show the predicted proton EDM; in particular, $|d_p| = 10^{-28}~e\,\mathrm{cm}$ on the solid line and $|d_p| = 10^{-29}~e\,\mathrm{cm}$, which is expected to be explored in the storage ring experiments, on the broken purple line.

The broad region of the $M_2$-$\mu_H$ plot will be able to be investigated by the storage ring experiment for the proton EDM.
It is expected that we can crosscheck the electron EDM and the nucleon EDMs in the broad parameter region in the near future.
In particular, $|d_p/d_e| \sim 10$ and $|d_n/d_e| \sim 15$ are predicted in the heavy higgsino system.
This is because the parameter dependences of the quark EDMs are similar to those of the electron EDM.
The difference of the quark EDMs from the electron one mainly arises from masses and charges of partons, QCD corrections below the electroweak scale, and the existence of $d_q^{ZH}$ contribution.

\section{Conclusions and Discussion \label{sec:conclusion}}

The electroweakly interacting fermions are promising candidates of new particles that appear as low-energy remnants in UV completions.
The EDMs are expected to be a good probe of $CP$ violation in the EWIMP fermions, and therefore we have focused on the EDM from the effective interactions among the Higgs doublet and the EWIMP fermions.
In particular, we have derived the anomalous dimensions for the dimension-five effective operators from the gauge interactions, the Yukawa interactions, and the Higgs quartic interaction.
The operators evolve through the anomalous dimensions from the input scale where the effective operators are generated to the electroweak scale.
Since $SU(2)_L$ is asymptotically free in the SM with a single electroweak interacting fermion, the quantum corrections tend to enhance the EDMs from the effective operators.

We have also evaluated the effect of the RGEs in specific models: the mini-split supersymmetric model with heavy higgsinos and several EWIMP models.
These models provide a candidate of the DM, and therefore they have been gathering attention as simplified models.

We have discussed the EDMs from the EWIMP fermions that contain an electromagnetically neutral state.
In \cref{sec:MDM}, we have especially considered the following fermions:
a $(n,Y_\chi)=(5,0)$ Majorana fermion,
a $(n,Y_\chi)=(3,0)$ Majorana fermion,
a $(n,Y_\chi)=(4,1/2)$ Dirac fermion, and
a $(n,Y_\chi)=(2,1/2)$ Dirac fermion.
Taking into account the RGE of the effective operators, we have discussed the impact of renormalization factors on the electron EDMs in each EWIMP DM scenario.
As results, we have found that the large correction from electroweak interactions arises if the effective operators consist of the large electroweak multiplet.

For the heavy higgsino split SUSY scenario, as we have discussed in \cref{sec:heavyhiggsino}, the correction from the gauge interactions can be of the order of $\Ocal(10)\%$.
Furthermore, it is expected that the future experiment for the electron EDM and nucleon EDMs will reach broad parameter space as shown in \cref{fig:eEDM_splitSUSY,fig:NEDM_splitSUSY}.
It is also expected that the wino mass region where we have studied will be covered using the disappearing track search at the 100~TeV $pp$ colliders \cite{Cirelli:2014dsa,CidVidal:2018eel}.

Finally, we comment on what we have not discussed in this work.
We have not considered the models that add two or more EWIMP fermions.
When we add two or more EWIMP fermions, there might be Yukawa interactions among the EWIMPs and the SM Higgs boson.
These interactions also affect the evolution of the effective operators.
However, these corrections highly depend on the models, and in that case we cannot estimate the enhancement factor without assuming the size of the Yukawa couplings.
We have focused only on one-loop corrections to the effective operators because the higher corrections, such as the two-loop RGEs and the one-loop matching conditions, are expected to be more suppressed by loop factors; they are expected to be much smaller than a percent level.

We have not discussed the scalar EWIMP models in this study.
The scalar EWIMP can also have the $CP$-odd interactions with the SM Higgs.
However, the interactions highly depend on the representation of the scalar EWIMPs.
We will discuss the EDMs in the scalar EWIMP models somewhere else.

\subsection*{Acknowledgement}
We would like to thank Eibun Senaha for valuable comments on this manuscript.
The work of T. K. is supported by IBS under the project code, IBS-R018-D1.
The work of R. N. is supported by KAKENHI Grant Numbers 16H06490, 18H05542, and 19K14701.
R. N. would like to thank Center for Theoretical Physics of the Universe, Institute for Basic Science for hospitality while this work was initiated.
\appendix

\bibliographystyle{./utphys}
\bibliography{ref}
\end{document}